\documentstyle[12pt]{article}
\begin{document}
\tolerance=5000
\def\be{\begin{equation}}
\def\ee{\end{equation}}
\def\bea{\begin{eqnarray}}
\def\eea{\end{eqnarray}}
\def\nn{\nonumber \\}
\def\beaa{\begin{eqnarray*}}
\def\eeaa{\end{eqnarray*}}
\def\cF{{\cal F}}
\def\det{{\rm det\,}}
\def\Tr{{\rm Tr\,}}
\def\e{{\rm e}}
\def\etal{{\it et al.}}
\def\erp2{{\rm e}^{2\rho}}
\def\erm2{{\rm e}^{-2\rho}}
\def\er4{{\rm e}^{4\rho}}
\def\etal{{\it et al.}}

\ 

\vskip -2cm

\ \hfill
\begin{minipage}{3.5cm}
NDA-FP-?? \\
July 1999 \\
\end{minipage}

\vfill

\begin{center}
{\Large\bf Strong Coupling Limit of ${\cal N}=2$ SCFT Free Energy 
and Higher Derivative AdS/CFT Correspondence}

\vfill

{\sc Shin'ichi NOJIRI}\footnote{\scriptsize 
e-mail: nojiri@cc.nda.ac.jp, snojiri@yukawa.kyoto-u.ac.jp} and
{\sc Sergei D. ODINTSOV$^{\spadesuit}$}\footnote{\scriptsize 
e-mail: odintsov@mail.tomsknet.ru}

\vfill

{\sl Department of Mathematics and Physics \\
National Defence Academy, 
Hashirimizu Yokosuka 239, JAPAN}

\ 

{\sl $\spadesuit$ 
Tomsk Pedagogical University, 634041 Tomsk, RUSSIA \\
}

\ 

\vfill

{\bf abstract}

\end{center}

We study the role of higher derivative terms (Riemann curvature 
squared ones) in thermodynamics of SCFTs via AdS/CFT correspondence.
Using IIB string effective action (d5 AdS gravity)
 with such HD terms deduced from heterotic string via duality we calculate
strong coupling limit of ${\cal N}=2$ SCFT 
free energy with the account of next to leading term in large $N$ expansion.
It is compared with perturbative result following from boundary QFT.
Considering modification of such action where HD terms form Weyl squared 
tensor we found (strong coupling limit) free energy in such theory. It is
interesting that leading and next to leading term of large $N$ expanded free
energy may differ only by factor 
3/4 if compare with perturbative result. Considering HD gravity as 
bosonic sector of some (compactified) HD supergravity we suggest 
new version of AdS/CFT conjecture and successfully test it on the level of
free energies for ${\cal N}=2,4$ SCFTs.

\newpage

\noindent
1. AdS/CFT correspondence \cite{1} (for an excellent review, see 
\cite{AGMOO}) may provide new insights to strong coupling 
regions of SUSY QFTs. One example of this sort is given by 
strong coupling limit of ${\cal N}=4$ super YM theory free energy 
\cite{GKP,GKT} which differs by factor $3/4$ with perturbative 
result (boundary QFT) in the leading order of $1/N$ 
expansion\footnote{Thermodynamics of ${\cal N}=4$ super YM theory 
in relation with AdS/CFT correspondence has been discussed in 
numerous works\cite{Wother,NO}.}. 
It is quite 
interesting to understand what happens in the next order of $1/N$ 
expansion. Clearly that such analysis should be related with 
higher derivatives (HD) terms on SG side. In more general framework 
HD terms may help in better understanding of AdS/CFT correspondence 
or even in formulation of new versions of bulk/boundary conjecture.

That is the purpose of present work to study the role of HD terms in bulk
action in AdS/CFT correspondence. In the next two sections we 
find strong coupling limit of ${\cal N}=2$ SCFT free energy and compare it 
with perturbative result up to 
the terms of  next to leading order. HD terms are chosen in 
Riemann curvature squared form or in Weyl tensor squared form.
The metric of effective five-dimensional gravity is BH in AdS background. 
In the last section we suggest new HD AdS/CFT conjecture. We show that it
works well
on the level of comparison of free energies for SCFTs (or trace anomalies) 
if bulk sector is described by some HD (super)gravity.

\noindent
2.The trace anomaly of  $d=4$, ${\cal N}=2$ and 
${\cal N}=4$ SCFTs (bulk side calculation) 
has been found up to next to 
leading order in the $1/N$ expansion in refs. \cite{BNG,NO1}.
Even in next to leading order term it coincides with QFT result 
(for gravity side derivation of trace anomaly, mainly in ${\cal N}=4$ 
super YM case, see also refs.\cite{NO2,CA}). 
The ${\cal N}=2$ theory with the gauge group $Sp(N)$ arises as 
the low-energy theory on the world volume on $N$ D3-branes 
sitting inside 8 D7-branes at an O7-plane \cite{Sen}. 
The string theory dual to this theory has been conjectured 
to be type IIB string theory on $AdS_5\times X^5$ where 
$X_5=S^5/Z_2$ \cite{FS}, whose low energy effective action 
is given by\footnote{
The conventions of curvatures are given by
\beaa
&& R=g^{\mu\nu}R_{\mu\nu} \ ,\quad 
R_{\mu\nu}= -\Gamma^\lambda_{\mu\lambda,\nu}
+ \Gamma^\lambda_{\mu\nu,\lambda}
- \Gamma^\eta_{\mu\lambda}\Gamma^\lambda_{\nu\eta}
+ \Gamma^\eta_{\mu\nu}\Gamma^\lambda_{\lambda\eta} \\
&& {R^\lambda}_{\mu\nu\kappa}= \Gamma^\lambda_{\mu\kappa,\nu}
- \Gamma^\lambda_{\mu\nu,\kappa}
+ \Gamma^\eta_{\mu\kappa}\Gamma^\lambda_{\nu\eta}
- \Gamma^\eta_{\mu\nu}\Gamma^\lambda_{\kappa\eta} \ ,\quad 
\Gamma^\eta_{\mu\lambda}={1 \over 2}g^{\eta\nu}\left(
g_{\mu\nu,\lambda} + g_{\lambda\nu,\mu} - g_{\mu\lambda,\nu} 
\right)\ .
\eeaa
} (see the corresponding derivation in ref.\cite{BNG}).
\be
\label{bng3} 
S=\int d^5x \sqrt{g}\left\{{N^2 \over 4\pi^2}
\left(R + 12 \right) 
+ {6N \over 24\cdot 16\pi^2}R_{\mu\nu\rho\sigma}
R^{\mu\nu\rho\sigma}\right\}\ .
\ee
The overall factor of the action is different by ${1 \over 2}$ 
from that of the action which corresponds to ${\cal N}=4$ 
$SU(N)$ gauge theory. The latter action is given by type IIB 
(compactified) string theory on $AdS_5\times S^5$. 
The factor ${1 \over 2}$ comes from the fact that the volume 
of $X^5=S^5/Z_2$ is half of $S^5$ due to $Z_2$.
Note that Riemann curvature squared term in the above bulk 
action is deduced from heterotic string via heterotic-type I 
duality \cite{Arkadii} (dilaton is assumed to be constant).

Then the equations of motion have the following form:
\be
\label{aiii}
0=-{c \over 2}g_{\zeta\xi}\left(
R_{\mu\nu\rho\sigma} R^{\mu\nu\rho\sigma}
+ {1 \over \kappa^2} R - \Lambda \right)
+ c R_{\zeta\mu\nu\rho} R_\xi^{\mu\nu\rho}
+ {1 \over \kappa^2} R_{\zeta\xi} 
-4 c D_\nu D_\mu R^{\mu\ \nu}_{\ \zeta\ \xi}\ .
\ee
Here 
\be
\label{bng4}
{1 \over \kappa^2}={N^2 \over 4\pi^2}\ , \quad
c={6N \over 24\cdot 16\pi^2}\ , \quad 
\Lambda=-{12 \over \kappa^2}=-{12N^2 \over 4\pi^2}\ .
\ee
We now treat the next-to-leading term of order $N$ as 
perturbation of order $N^2$ terms. In the leading order, a 
solution is given by 
\be
\label{sp1b}
ds^2=g_{\mu\nu}dx^\mu dx^\nu
=-\e^{2\rho}dt^2+\e^{-2\rho}dr^2 + r^2\sum_{i=1}^3\left(
dx^i \right)^2 \ ,\ \ 
\e^{2\rho}={1 \over r^2}\left(-\mu + r^4 \right)\ .
\ee
In the metric (\ref{sp1b}), we find
\bea
\label{Rie}
&& D_\nu D_\mu {{{R^\mu}_\zeta}^\nu}_\xi=0\ ,\quad 
R_{\mu\nu\rho\sigma}R^{\mu\nu\rho\sigma}
=\left( 40 + {72 \mu^2 \over r^8}\right) \nn \ .
\eea
Then the metric is modified by 
\bea
\label{sp1bb}
&& ds^2=g_{\mu\nu}dx^\mu dx^\nu
=-\e^{2\rho}dt^2+\e^{-2\rho}dr^2 + r^2\sum_{i=1}^3\left(
dx^i \right)^2 \nn
&& \e^{2\rho}={1 \over r^2}\left\{-\mu + \left(1+{2 \over 3}\epsilon 
\right)r^4 +2\epsilon {\mu^2 \over r^4} \right\}\ , \quad 
\epsilon \equiv c\kappa^2 = {1 \over 16N}\ .
\eea
Then the radius $r_h$ of the horizon and the temperature $T$ 
are given by
\be
\label{sp3}
r_h\equiv \mu^{1 \over 4}\left(1 - {2 \over 3}\epsilon \right)\ ,
\quad T={\mu^{1 \over 4} \over \pi }\left(1 - 2\epsilon\right)
={\mu^{1 \over 4} \over \pi }\left(1 - {1 \over 8N}\right)\ .
\ee

We now consider the thermodynamical quantities like free energy.
After Wick-rotating the time variables by $t\rightarrow i\tau$, 
the free energy $F$ can be obtained from the action $S$ in 
(\ref{bng3}) where the classical solution is substituted:
\be
\label{F1}
F={1 \over T}S\ .
\ee
Using (\ref{aiii}), (\ref{bng4}) and (\ref{sp1bb}), we find
\bea
\label{sp5}
S&=&{N^2 \over 4\pi^2}\int d^5x \sqrt g \left\{ 8 
- {2\epsilon \over 3}\left(40 + {72 \mu^2 \over r^8}
\right)\right\} \nn
&=& {N^2V_3 \over 4\pi^2 T}\int_{r_h}^\infty dr r^3 
\left\{ 8 - {2\epsilon \over 3}\left(40 + {72 \mu^2 \over r^8}
\right)\right\}\ .
\eea
Here $V_3$ is the volume of 3d flat space and we assume 
$\tau$ has a period of ${1 \over T}$. The expression of $S$ 
contains the divergence coming from large $r$. In order to 
subtract the divergence, we 
regularize $S$ in (\ref{sp5}) by cutting off the integral at 
a large radius $r_{\rm max}$ and subtracting the solution 
with $\mu=0$:
\bea
\label{sp7}
S_{\rm reg}&=&{N^2V_3 \over 4\pi^2 T}\left(\int_{r_h}^\infty dr r^3 
\left\{ 8 - {2\epsilon \over 3}\left(40 + {72 \mu^2 \over r^8}
\right)\right\}\right. \nn 
&& \left. - \e^{\rho(r=r_{\rm max}) - \rho(r=r_{\rm max};\mu=0)}
\int_0^{r_{\rm max}} dr r^3\right)
\left\{ 8 - {80\epsilon \over 3}\right\}\ .
\eea
The factor $\e^{\rho(r=r_{\rm max}) - \rho(r=r_{\rm max};\mu=0)}$ 
is chosen so that the proper 
length of the circle which corresponds to the period ${1 \over T}$ 
in the Euclidean time at $r=r_{max}$ coincides 
with each other in the two solutions. Then we find
\be
\label{sp8}
F=-{N^2V_3\left(\pi T\right)^4 \over 4\pi^2 }\left(1 + {3 \over 4N}\right)\ .
\ee
The entropy ${\cal S}$ and the mass (energy) $E$ are given 
by
\bea
\label{sp9}
{\cal S}&=&-{dF \over dT}={N^2V_3\left(\pi T\right)^4  \over \pi^2 T }
\left(1 + {3 \over 4N}\right) \nn
E&=&F+T{\cal S}={3N^2V_3\left(\pi T\right)^4  \over 4\pi^2 }
\left(1 + {3 \over 4N}\right)\ .
\eea

We now compare the above results with those of field theory of 
${\cal N}=2$ $Sp(N)$ gauge theory. ${\cal N}=2$ theory contains 
$n_V=2N^2+N$ vector multiplet and $n_H=2N^2+7N-1$ hypermultiplet 
\cite{BNG}. Vector multiplet consists of two Weyl fermions, one 
complex scalar and one real vector what gives 4 bosonic (fermionic) 
degrees of freedom on shell and hypermultiplet contains two complex 
scalars and two Weyl fermions, what also gives 4 bosonic (fermionic) 
degrees of freedom on shell \cite{SW}. Therefore there appear 
$4\times \left(n_V + n_H\right)=16 \left(N^2 + 2N 
-{1 \over 4}\right)$ boson-fermion pairs. 
In the limit which we consider, the interaction between 
the particles can be neglected. The contribution to the free 
energy from one boson-fermion pair in the space with the 
volume $V_3$ can be easily estimated \cite{GKP,GKT}. 
Each pair gives a contribution to the free energy of 
${\pi^2 V_3T^4 \over 48}$. Therefore the total free energy 
$F$ should be 
\be
\label{ff1}
F=-{\pi^2 V_3 N^2 T^4 \over 3}\left(1+{2 \over N}-{1 \over 4N^2}
\right)\ .
\ee
Comparing (\ref{ff1}) with (\ref{sp8}), there is the difference 
of factor ${4 \over 3}$ in the leading order of $1/N$ as observed 
in \cite{GKP,GKT}. 

Hence, we calculated strong coupling limit of 
free energy in  ${\cal N}=2$ SCFT from 
SG side up to next to leading order term (it was generated by 
Riemann curvature squared term). Its weak coupling limit 
(\ref{ff1}) obtained from QFT side cannot be presented as 
strong coupling limit free energy multiplied to some constant. 
This only holds for leading order terms where mismatch multiplier is 
3/4.  The next to leading term in (\ref{ff1}) should be multiplied to 
$9/32$ in order to produce the corresponding term in (\ref{sp8}). 
We have to pay attention once more that the main role in whole 
above analysis was played by Riemann curvature squared term. 
In the  section 4 we try to understand the role of HD  
terms in AdS/CFT correspondence from the different point 
of view. 

\noindent
3. We now consider the case where the action is given by the 
sum of Einstein term and the square of the Weyl tensor 
$C_{\mu\nu\rho\sigma}C^{\mu\nu\rho\sigma}$\footnote{
We thank A. Tseytlin for suggestion to write this section. 
His motivation was that the case with $R^2$ combination as $C^2$ 
does not make modification of original 
${\rm AdS}_5\times{\rm S}_5$ solution as it already happened 
with $C^4$ correction \cite{GKT}.
}
\be
\label{W1}
S=-\int d^5 x \sqrt{- G}\left\{ {1 \over \kappa^2} R - \Lambda 
+ \tilde c C_{\mu\nu\rho\sigma}C^{\mu\nu\rho\sigma} \right\} \ .
\ee
In five dimensions, the square of the Weyl tensor 
is given by
\be
\label{W2}
C_{\mu\nu\rho\sigma}C^{\mu\nu\rho\sigma}
= {1 \over 6} R^2 - {4 \over 3} R_{\mu\nu} R^{\mu\nu}
+ R_{\mu\nu\rho\sigma} R^{\mu\nu\rho\sigma}\ .
\ee
The Weyl tensor term 
does not contribute to the radius $L$ of ${\rm AdS}_5$. 
By using the equation of motion, we find
\be
\label{a1}
0=-{3 \over 2\kappa^2}R - {1 \over 2}\tilde c
C_{\mu\nu\rho\sigma}C^{\mu\nu\rho\sigma} + {5 \over 2}\Lambda\ .
\ee
By using (\ref{a1}), we can delete $R$ in (\ref{W1}) and obtain 
\be
\label{a2}
S=-\int d^5 x \sqrt{- G}\left\{ {2 \over 3}  \Lambda + {2 \over 3}
\tilde c C_{\mu\nu\rho\sigma}C^{\mu\nu\rho\sigma} \right\} \ .
\ee
Since the square of the Weyl tensor is given by 
\be
\label{a3}
C_{\mu\nu\rho\sigma}C^{\mu\nu\rho\sigma} 
= {72 \mu^2 \over r^8}
\ee
for the leading solution, which is given by
\be
\label{a4}
\e^{2\rho}={\mu \over r^2}\left({r^4 \over r_0^4} - 1\right)\ ,
\quad r_0^4 \equiv - {12\mu \over \kappa^2 \Lambda}\ ,
\ee
we find 
\be
\label{a5}
S=\int d^5x \sqrt g 
\left\{ {2 \over 3}\Lambda + 48\tilde c {\mu^2 \over r^8}\right\}\ .
\ee
The horizon radius and the temperature are given by
\be
\label{sp3b}
r_h\equiv r_0\left(1 - {1 \over 2}\epsilon \right)\ ,
\quad T=-{\kappa^2 \Lambda r_0 \over 12\pi }
\left(1 - {5 \over 2}\epsilon\right) \ , \quad 
\epsilon \equiv -{\tilde c\Lambda \kappa^4 \over 12} \ .
\ee
Then by regularizing the action (\ref{a5}) as in (\ref{sp7}), 
we find
\bea
\label{a6}
S_{\rm reg}&=&{N^2V_3 \over 4\pi^2 T}\left(\int_{r_h}^\infty dr r^3 
\left\{ {2 \over 3}\Lambda + 48\tilde c {\mu^2 \over r^8}
\right\}\right. \nn 
&& \left. - \e^{\rho(r=r_{\rm max}) - \rho(r=r_{\rm max};\mu=0)}
\int_0^{r_{\rm max}} dr r^3\left({2 \over 3}\Lambda\right) 
\right) \nn
&& ={V_3 \over T}\left( - {\Lambda \over 12}\right)^{-3}
{\left(\pi T\right)^4 \over \kappa^8}
\left(1 + 18 \epsilon + {\cal O}\left(\epsilon^2\right) 
\right) \
\eea
Remarkably the above result coincides with that of the simplified 
procedure, where the correction is evaluated by substituting the 
leading solution into the correction term, the square of the Weyl 
tensor, in the action (\ref{W1}):
\bea
\label{c1}
S_{W^2}&=&- \tilde c \int d^5 x \sqrt{- G}
C_{\mu\nu\rho\sigma}C^{\mu\nu\rho\sigma} \nn
&=&{N^2V_3 \over 4\pi^2 T}\int_{r_0}^\infty dr r^3 
\left( 72\tilde c {\mu^2 \over r^8} \right) \nn
&=&{V_3 \over T}\left( - {\Lambda \over 12}\right)^{-3}
{18 \epsilon \left(\pi T\right)^4 \over \kappa^8}
+ {\cal O}\left(\epsilon^2\right) \ .
\eea
Therefore we obtain the following thermodynamical quantities;
\bea
\label{W6}
&& F=- V_3 \left( - {\Lambda \over 12}\right)^{-3}
{\left(\pi T\right)^4 \over \kappa^8}\left(
1+18\epsilon \right) \ ,\quad 
{\cal S}=4V_3 \left( - {\Lambda \over 12}\right)^{-3}
{\left(\pi T\right)^4 \over T\kappa^8}\left(1+18\epsilon \right) 
\ ,\nn
&& \quad E=3V_3 \left( - {\Lambda \over 12}\right)^{-3}
{\left(\pi T\right)^4 \over \kappa^8}\left(1+18\epsilon \right)  \ .
\eea
Choosing $\kappa$ and $\Lambda$ as in (\ref{bng4}),
above action may be considered as another modification of 
string action of the same sort as (\ref{bng3}).
With it we obtain
\bea
\label{W6b}
&& F=-{N^2 V_3 \left(\pi T\right)^4 \over 4\pi^2}\left(1
+{18\cdot 4\pi^2 \tilde c \over N^2} \right) \ ,\quad 
{\cal S}={4 N^2 V_3 \left(\pi T\right)^4 \over 4\pi^2 T }\left(1+
{18\cdot 4\pi^2 \tilde c \over N^2} \right) \ ,\nn
&&  E={3 N^2 V_3 \left(\pi T\right)^4 \over 4\pi^2}\left(1+
{18\cdot 4\pi^2 \tilde c \over N^2} \right) \ .
\eea
Therefore if we further choose $\tilde c$ by\footnote{
Note that such choice of $\tilde c$ is not justified 
by string considerations. Actually, string theory indicates that 
$\tilde c$ should coincide with $c$ from Eq.(\ref{bng3}). 
However, there are arguments that additional stringy 
corrections (for example, from antisymmetric tensor 
field) may appear.}
\be
\label{W6c}
\tilde c={N \over 9\cdot 4\pi^2}\ ,
\ee
the strong coupling limit free energy in (\ref{ff1}) can be 
reproduced including the 
next-to-leading order term with the common overall 
factor ${3 \over 4}$.

\noindent
4. Instead of (\ref{bng3}), we can consider more general action of 
$R^2$ gravity:
\be
\label{vi}
S=-\int d^5 x \sqrt{- G}\left\{a R^2 
+ b R_{\mu\nu} R^{\mu\nu}
+ c R_{\mu\nu\rho\sigma} R^{\mu\nu\rho\sigma}
+ {1 \over \kappa^2} R - \Lambda \right\} \ .
\ee
One can imagine that this is bosonic sector of some HD 
multidimensional (probably compactified to AdS) SG. 
It was already shown \cite{NO1} that such bulk theory 
may correctly reproduce trace anomaly of ${\cal N}=4$ 
super YM theory. We wish further check such HD AdS/CFT 
conjecture on the level of free energies. Note that such theory 
may represent (yet unknown) resummation of string effective 
action or it may directly follow from strings or M-theory. 
It is known, for example, 
that HD quantum gravity has better UV properties than usual 
Einstein gravity (see \cite{BOS} for a review).

If $c=0$, the equations of motion given by (\ref{vi}) can be 
solved exactly. And as shown in \cite{NO1}, the anomaly of 
${\cal N}=4$ super Yang-Mills theory can be reproduced in 
case of $c=0$. In the following, we consider only special 
case:$c=0$. 

When $c=0$, we can assume that the solution has the form: 
\be
\label{sp1}
ds^2=g_{\mu\nu}dx^\mu dx^\nu
=-\e^{2\rho}dt^2+\e^{-2\rho}dr^2 + r^2\sum_{i=1}^3\left(
dx^i \right)^2 \ ,\ \ 
\e^{2\rho}={1 \over r^2}\left(-\mu + {r^4 \over L^2}\right)\ .
\ee
In this case, the curvature tensors become
\be
\label{aii}
R=-{ 20 \over L^2}\ ,\ \ 
R_{\mu\nu}=-{4 \over L^2}g_{\mu\nu}\ ,
\ee
which tell that these curvatures are covariantly constant. 
Then in the equations of motion following from 
the action (\ref{vi}), the terms containing the covariant 
derivatives of the curvatures 
vanish and the equations have the following forms:
\be
\label{aiiib}
0=-{1 \over 2}g_{\zeta\xi}
\left\{a R^2 + b R_{\mu\nu} R^{\mu\nu}
+ {1 \over \kappa^2} R - \Lambda \right\} 
+ 2a R R_{\zeta\xi} 
+ 2b R_{\mu\zeta}{ R^\mu}_\xi
+ {1 \over \kappa^2} R_{\zeta\xi}\ .
\ee
Then  substituting Eqs.(\ref{aii}) into (\ref{aiiib}), we find 
\be
\label{aivb}
0={80a \over L^4} + {16b \over L^4} 
- {12 \over \kappa^2 L^2} -\Lambda\ .
\ee
The equation (\ref{aivb}) can be solved with respect to $L^2$ 
if 
\be
\label{aiva}
{144 \over \kappa^4}-16\left\{20a + 4b \right\}
\Lambda\geq 0
\ee
which can been found from the determinant in (\ref{aivb}).
Then we obtain
\be
\label{ll}
L^2=-{{12 \over \kappa^2}\pm \sqrt{
{144 \over \kappa^4}-16 \left\{20a + 4b \right\}
\Lambda} \over 2\Lambda}\ .
\ee
The sign in front of the root in the above equation 
may be chosen to be positive, which 
corresponds to the Einstein gravity ($a=b=0$). 
With $L$ in (\ref{ll}), the horizon radius $r_h$ and 
the temperature $T$ are given by 
\be
\label{sp3c}
r_h\equiv \mu^{1 \over 4}L^{1 \over 2}\ ,\quad 
T={\mu^{1 \over 4} \over \pi L^{3 \over 2}}\ .
\ee
and we can find 
the free energy $F$, the entropy ${\cal S}$ and the energy $E$
by generalizing (\ref{sp8}), (\ref{sp9}):
\bea
\label{rscr1}
&& F=-{\mu V_3\over 8}\left({8 \over \kappa^2} - {320 a \over L^2} 
- {64 b \over L^2} \right) \ ,\quad 
{\cal S}={\mu V_3 \over 2T}\left({8 \over \kappa^2} 
- {320 a \over L^2} - {64 b \over L^2} \right) \nn
&& E={3\mu V_3 \over 8}\left({8 \over \kappa^2} - {320 a \over L^2} 
- {64 b \over L^2} \right) \ .
\eea
Here we delete $\Lambda$ by using (\ref{ll}). Note that 
we may also consider special case of no Einstein term 
in above equations since the above expressions are not perturbative 
but exact when $c=0$.

We can consider the case of ${\cal N}=4$ super Yang-Mills theory. 
When the gauge group is $U(N)$, there are $8N^2$ set of the 
bosonic and fermionic degrees of freedom on-shell and when 
$SU(N)$, $8(N^2 -1)$ since one multiplet corresponds to $n_V=n_H=1$ 
in ${\cal N}=2$ theory and contains 8 boson (fermion) degrees of 
freedom on shell. Then from the perturbative QFT its free energy 
is given by, 
\be
\label{N4}
F= \left\{\begin{array}{ll}
-{\pi^2 V_3 N^2 T^4 \over 6} \quad &U(N)\ \mbox{case} \\
-{\pi^2 V_3 N^2 T^4 \over 6}\left(1 -{1 \over N^2}\right) &SU(N)\ 
\mbox{case} \\ \end{array}\right. \ .
\ee
If we assume $\kappa^2$, $\Lambda$, $a$, and $b$ are given 
by the powers of $N$, we find the above free energy $F$ can be 
reproduced if
\be
\label{N4ii}
\kappa^2={6\pi^2 \over N^2}\ ,\quad 
\Lambda=-{2N^2 \over \pi^2} \ ,\quad
10a+2b= \left\{ \begin{array}{ll} 
0 &U(N)\ \mbox{case} \\
{1 \over 24\pi^2} \quad &SU(N)\ \mbox{case} \\
\end{array} \right. \ . 
\ee
As a special case, we can consider the theory without Einstein 
term, i.e., ${1 \over \kappa^2}=0$. Then we find $L^2$ 
in (\ref{ll}) is given by
\be
\label{llii}
L^2=\pm 2\sqrt{20a+4b \over -\Lambda}\ .
\ee
Since the $-$ sign in (\ref{llii}) does not correspond to 
AdS but to de Sitter space, we only consider the case of 
$+$ sign. Then the free energy $F$ in (\ref{rscr1}) has the 
following form:
\be
\label{N4b}
F=\pm\mu V_3\sqrt{-\Lambda\left(20a+4b \right)} 
=-{4V_3\left(\pi T\right)^4 \left(10a+2b \right)^2 
\over \Lambda} \ .
\ee
In the $\pm$ sign in the first line, $+$ corresponds to negative 
$\Lambda$ and $-$ to positive one since $20a+4b>0$ if 
$\Lambda<0$ and $20a+4b<0$ if $\Lambda>0$ from (\ref{aiva}). 
Therefore we can obtain the free 
energy in (\ref{N4}) if
\be
\label{N4iii}
{\left(10a+2b \right)^2 \over \Lambda}=
\left\{ \begin{array}{ll}
{N^2 \over 24\pi^2} \quad &U(N)\ \mbox{case} \\
{N^2 - 1 \over 24\pi^2} \quad &SU(N)\ \mbox{case} \\
\end{array} \right.\ . 
\ee
Note that $\Lambda$ should be positive. 
In \cite{NO1}, it has been shown that the conformal anomaly 
of ${\cal N}=4$ super Yang-Mills theory can be reproduced by 
above $R^2$ gravity if
\be
\label{N4CA}
c=0\ ,\quad {L^3 \over \kappa^2} - 40aL -8b L 
= {2N^2 \over \left(4\pi\right)^2}\ .
\ee
It is remarkable  that (\ref{N4CA}) can be compatible 
with (\ref{N4iii}) if 
\be
\label{N4iv}
{1 \over \kappa^2}=c=0\ ,\quad 
10a+2b = - {3^{1 \over 3} \over 2^{8 \over 3}}\cdot 
{N^2 \over \left(4\pi\right)^2}\ , \quad
\Lambda = - {3^{2 \over 3} \over 2^{16 \over 3}}\cdot 
{N^2 \over \left(4\pi\right)^2}\ ,
\ee
in the leading order of ${1 \over N}$ expansion, or 
$U(N)$ case. 

\noindent
5. In summary, we calculated strong coupling limit for free energy of ${\cal
N}=2$ SCFT  from AdS/CFT correspondence in the next to leading order 
of $1/N$ expansion. As bulk side we used AdS Einstein gravity with 
Riemann curvature (or Weyl tensor) squared term. For general
five-dimensional HD gravity considered as bosonic sector of some HD 
(probably compactified) SG we formulated new HD AdS/CFT conjecture 
which works on the level of free energies for ${\cal N}=4$ super YM. 
Note also that it works for ${\cal N}=2$ SCFT. Indeed,
in this case the analog of eqs.(\ref{N4ii}) is given by
\be
\label{N2a}
\kappa^2={3\pi^2 \over N^2}\ ,\quad 
\Lambda=-{4N^2 \over \pi^2}\ ,\quad 
 10a+2b=-{N \over 9\pi}+{\cal O}(1)\ .
\ee
Notice finally that it would be really interesting to consider 
the role of above HD terms to thermodynamics of AdS/CFT 
correspondence in the case of non-constant dilaton. Without HD 
terms the corresponding (approximate) 
AdS BH solution has been found in ref.\cite{NO}. 

It could be of interest also to investigate the role of spatial 
curvature to above analysis. Without HD terms such investigation 
has been presented in refs.\cite{BCM}, which is much related to the 
study of finite gauge theories, including ${\cal N}=4$ super YM 
theory in curved spacetime (QFT side) where curvature squared and 
scalar-gravitational divergences appear, see refs.\cite{BLO}. 

\noindent
{\bf Acknoweledgements.} We are very grateful to A.A. Tseytlin for 
participation at early stages of this work and many helpful 
discussions.

\end{document}